\documentclass[twocolumn,secnumarabic,amssymb, nobibnotes, aps, prd]{revtex4}
\usepackage{graphicx}

\begin{document}
\title{Observation of flux qubit states with the help of a superconducting differential double contour interferometer}
\author{A.V. Nikulov}
\affiliation{Institute of Microelectronics Technology and High Purity Materials, Russian Academy of Sciences, 142432 Chernogolovka, Moscow District, RUSSIA.} 
\begin{abstract} The quantum states of flux qubit is suggested to observe with the help of a new device, the superconducting differential double contour interferometer (DDCI). The flux qubit and the superconducting quantum interference device (DC-SQUID) are connected in the DDCI through the phase of the wave function rather than through magnetic flux. The critical current of DC-SQUID should change to the maximum value at the change of the flux qubit state thanks to this phase coupling. A large jump in the critical current and voltage enables to observe continuously the change in time the state of the flux qubit. This observation can have fundamental importance for the investigation of the superposition of macroscopic quantum states. 
\end{abstract}

\maketitle  

\narrowtext

\section{Introduction}
Superconducting flat loop interrupted by one or more Josephson junctions is of great interest both in the context of the extrapolation of the predictions of quantum mechanics towards the macroscopic level \cite{Leggett1985,LGineq2014,LGineq2016} and as qubit - possible elements in a future quantum computer \cite{Clarke2008,Devoret2013,Wendin2017,China2018}. Such loop is known now as the persistent-current qubit \cite{Mooij1999} or flux qubit \cite{Mooij2003,Clarke2003}. The flux qubit interrupted by three Josephson junctions is investigated in the most works \cite{Mooij1999,Mooij2003,Clarke2003,Semba2017}. The superconducting current through the Josephson junctions $I_{s1} = I_{1} \sin \varphi_{1} $, $I_{s2} = I_{2} \sin \varphi_{2} $, $I_{s3} = I_{3} \sin \varphi_{3} $ is determined by their critical current $I_{1}$, $I_{2}$, $I_{3}$ and the phase difference between their boundaries $\varphi_{1}$, $\varphi_{2}$, $\varphi_{3}$. In the stationary state, the currents through the junctions in the closed loop must be equal $I_{s1} = I_{s2} = I_{s3} = I_{p}$. 

The phase differences are determined by the equation $\varphi_{1} + \varphi_{12} + \varphi_{2} + \varphi_{23} + \varphi_{3} + \varphi_{31} = 2\pi n_{q}$, which is derived from the requirement $\oint_{l}dl \nabla \varphi = 2\pi n_{q}$ that the complex wave function must be single-valued $\Psi = |\Psi |e^{i\varphi } =  |\Psi |e^{i(\varphi + n2\pi )} $ in any point of the loop. The phase change $\varphi_{12}$, $\varphi_{23}$, $\varphi_{31}$ along segments of the loop between Josephson junctions is determined by the canonical momentum of Cooper pairs $p = (\Psi ^{*}\hat{p}\Psi )/n_{s} = -i\hbar(\Psi ^{*}\nabla \Psi )/n_{s} = \hbar \nabla \varphi $, which $p = mv + qA$ depends on the velocity Cooper pairs $v$ and the vector potential $A$. The velocity $v = I_{p}/sqn_{s}$ is neglected, since the cross section $s$ and the density of Cooper pairs $n_{s} = |\Psi |^{2}$ in the loop segments are great. Therefore $\varphi_{12} + \varphi_{23} + \varphi_{31} \approx  q\Phi /\hbar = 2\pi\Phi/\Phi_{0}$ and $\varphi_{1} + \varphi_{2} + \varphi_{3} \approx 2\pi (n_{q}-\Phi/\Phi_{0})$, where $\Phi _{0} = 2\pi \hbar /q = \pi \hbar /e$ is the flux quantum. Thus, the persistent current $I_{p} = I_{1} \sin \varphi_{1} = I_{2} \sin \varphi_{2} = I_{3} \sin \varphi_{3}$ is determined by the magnetic flux $\Phi$ inside the loop and an integer quantum number $n_{q}$ determining the state of the flux qubit. For example 
$$I_{p} \approx I_{c,J}\sin \frac{2\pi }{3} (n_{q}-\frac{\Phi }{\Phi_{0}}) \eqno{(1)}$$
when the critical currents through the Josephson junctions are equal $I_{1} = I_{2} = I_{3} = I_{c,J}$. The two states $n_{q} = n'$ and $n_{q} = n'+1$ having a minimum and oppositely directed current $I_{p} \approx I_{c,J} \sin (\pi /3 + \delta \Phi /\Phi_{0})$ and $I_{p} \approx -I_{c,J}\sin (\pi /3 - \delta \Phi /\Phi_{0})$ are considered in the theory of the flux qubit at $\Phi = (n'+0.5)\Phi_{0} + \delta \Phi $, where $\delta \Phi \ll \Phi_{0}$.

\section{Superposition of macroscopic quantum states}
The superconducting loop interrupted by the Josephson junctions is a qubit if  superposition of states $n_{q} = n'$ and $n_{q} = n'+1$ can be assumed. This superposition is considered as an analogue of the superposition of spin 1/2 projections \cite{Leggett2003} and is written in the same way \cite{Clarke2008}
$$\psi  = \alpha |\uparrow  > +  \beta |\downarrow  > \eqno{(2)}$$
The Hamiltonian of the flux qubit 
$$H_{q} = \epsilon \sigma _{z} - \Delta \sigma _{x} \eqno{(3)}$$
is also written \cite{Mooij2003,Leggett2014} by analogy with the Hamiltonian of spin 1/2 using Pauli matrices $\sigma _{z}$, $\sigma _{x}$ \cite{LL}. Energy of the states $n_{q} = n'$ ($|\uparrow >$) and $n_{q} = n ' +1$ ($|\downarrow > $) \cite{Leggett2014}
$$\epsilon = |I_{p}\delta \Phi | \eqno {(4)}$$
depends on the deviation $\delta \Phi$ of the magnetic flux $\Phi = (n'+0.5)\Phi_{0} + \delta \Phi = (n'+0.5 + \delta f)\Phi_{0}$ from $\Phi = (n'+0.5)\Phi_{0}$. For $\delta \Phi = 0$, the energies of the two states are equal $\epsilon = 0$. But the energy levels should be split in the presence of quantum tunneling with tunneling energy $ \Delta $ \cite{Clarke2008,Clarke2003}. The observation of the splitting of the levels \cite{Mooij2003}, the Rabi oscillations \cite{Mooij2003} and some other effects are considered to be the experimental evidences of superposition of states $n_{q} = n'$ ($|\uparrow >$) and $n_{q} = n'+1$ ($|\downarrow > $) since these effects are consistent with a textbook quantum mechanical prediction, which generally ascribes a non-zero complex amplitude to each of the states (1). But the Rabi oscillations, as it is noted correctly in \cite{LGineq2016}, is not necessarily inconsistent with a classical 'value-definite' description, which prescribes that the system is in exactly one state at any given moment. The splitting of the levels and other effects also does not guarantee evidence of superposition of states since the assumption of superposition of macroscopic quantum states of flux qubit contradicts to macroscopic realism \cite{Leggett1985}. Special theorems (so-called no-go or no-hidden-variables theorems \cite{Mermin1993}) were proposed to prove the impossibility of realistic description (i.e. without the assumption of superposition of states) quantum phenomena. A.J. Leggett and A. Garg have suggested such theorem for macroscopic quantum systems, considering as example the superconducting loop interrupted by the Josephson junctions \cite{Leggett1985}. 

\section{Observation of the eigenstates of flux qubit}
According to the formalism of quantum mechanics formulated by von Neumann \cite{Neumann1932} quantum state should change in two fundamentally different way: Process 1 - the discontinuous change at measurement, in which the jump from an original state $\psi = \Sigma _{i}a_{i}\phi _{i}$ to an eigenstate $\phi _{i}$ of a dynamical variable that is being observed (with the probability $|a_{i}|^{2}$) occurs and Process 2 - the continuous deterministic change of the state $\psi(t)$ or $\phi _{i}(t)$ of an isolated system with time. The jump during measurement was postulated first by Dirac in 1930 \cite{Dirac1930}. Therefore it is called sometimes the Dirac jump. Having made the statement "{\it after the first measurement has been made, there is no indeterminacy in the result of the second}" \cite{Dirac1930} Dirac postulated a change in the quantum state: "{\it In this way we see that a measurement always causes the system to jump into an eigenstate of the dynamical variable that is being measured}" \cite{Dirac1930}. Dirac jump is known also as wave function collapse or reduction of quantum state, in terms introduced by von Neumann \cite{Neumann1932}. Thus, the measurement process plays an active role in quantum mechanics. When measuring, the quantum system must jump from the superposition of states (1) to the eigenstate of the dynamical variable that is being measured: in the eigenstate $|\uparrow >$ with the probability $|\alpha |^{2}$ or in the eigenstate $|\downarrow > $ with the probability $|\beta |^{2}$. Therefore it is important how the states of the flux qubit are measured \cite{Leggett2014}.  

In most experiments to date this has been done by coupling the flux qubit inductively to a dc SQUID (Superconducting Quantum Interference Device, a superconducting loop interrupted by two Josephson junction) \cite{Mooij2000,Tanaka2002,Tanaka2009}. The critical (switching) current of the dc SQUID depends periodically $I_{c}(\Phi )$ on the total magnetic flux $\Phi $ threads the dc SQUID loop. The total magnetic flux is the sum $\Phi = \Phi_{ext} + \Delta \Phi _{I} = BS + LI_{p}$ of the great flux $\Phi_{ext} = BS$ of external magnetic field $B$ and the small magnetic flux $\Delta \Phi _{I} = LI_{p}$ induced by the persistent current $I_{p}$ (1) of the flux qubit \cite{Mooij2000,Tanaka2002,Tanaka2009}. The fundamental quantity,  whose behaviour is of interest, is the small magnetic flux $\Delta \Phi _{I} = LI_{p}$. This quantity changes on the value $2LI_{c,J}\sin \pi /3 $ with the jumps of the state of flux qubit between $n_{q} = n'$ and $n_{q} = n'+1$. The value $2LI_{c,J}\sin \pi /3 $ is usually much less (about a hundred times) the great flux $\Phi_{ext} = BS$ of external magnetic field $B$ \cite{Mooij2000,Tanaka2002,Tanaka2009}. Thus,  small changes of the critical current of the dc SQUID $I_{c}(BS + LI_{p})$ induced by the jump of the flux qubit state is observed against the background of a large change connected the external magnetic field $B$ \cite{Mooij2000,Tanaka2002,Tanaka2009}. Therefore method of the measurements of the small magnetic flux $\Delta \Phi _{I} = LI_{p}$ differs from the common method of measuring the magnetic flux with the help the dc SQUID \cite{SQUIDs1977,Barone1982}. 

The voltage at a bias current $I_{b}$ rather than the switching current is measured  usually \cite{SQUIDs1977,Barone1982}. The voltage depends on the critical current  of the dc SQUID $I_{c}(\Phi )$ when $I_{b} > I_{c}$  and therefore the voltage change allows to detect the change of the magnetic flux $\Phi $ \cite{SQUIDs1977,Barone1982}. This method is difficult to use in the case of the flux qubit because the change in the critical current of the dc SQUID is very small when the state changes between $n_{q} = n'$ and $n_{q} = n'+1$. Therefore the switching current $I_{sw}$ is measured \cite{Mooij2000,Tanaka2002,Tanaka2009}. In order to measure the switching current $I_{sw}$ the bias current increases from $I_{b} = 0$ to $I_{b} \approx  I_{c}(\Phi )$ when the escape of the dc SQUID from the zero-voltage state is observed. The states of the flux qubit influence on the value of the switching current due to its inductive coupling with the dcSQUID \cite{Mooij2000,Tanaka2002,Tanaka2009}. 

Each run of the bias current from $I_{b} = 0$ to $I_{b} \approx  I_{c}(\Phi )$ may be considered as an act of measurement of the flux qubit state. The authors \cite{Tanaka2002,Tanaka2009} repeat the act of measurement tens of thousands of times in a narrow region of the magnetic flux $0.03 > \delta \Phi > -0.03$ and observe the values of the dc SQUID switching current corresponding to the states $n_{q} = n'$ and $n_{q} = n'+1$ of the flux qubit. Such a measurement protocol is very far from the von Neumann projective scheme \cite{Leggett2014}. Moreover the same value of the switching current should be observed since the same dynamical variable is measured and the system should jump into its eigenstate at the first measurement according to the Dirac postulate \cite{Dirac1930}. The different states $n_{q} = n'$ or $n_{q} = n'+1$ of the superposition (2) are observed \cite{Tanaka2002,Tanaka2009} contrary to the Dirac postulate \cite{Dirac1930}. The probability $P$ of finding the states $n_{q} = n'$ or $n_{q} = n'+1$ depends on the energy (4) and is describe by the Arrhenius law  
$$P(n') \approx \frac{\exp -\epsilon/k_{B}T}{\exp -\epsilon/k_{B}T + \exp \epsilon/k_{B}T} = \frac{1}{1 + \exp 2\theta \delta f} \eqno {(5)}$$ 
according to the measurement results \cite{Tanaka2004}. Here $\theta = I_{p}\Phi_{0}/k_{B}T$. The contradictions between the measurement results \cite{Tanaka2002,Tanaka2004,Tanaka2009} and the basics of quantum mechanics makes it relevant alternative methods for observation the states of the flux qubit. 

\section{The superconducting differential double contour interferometer (DDCI)}
The method proposed in this work is based on a new device, the superconducting differential double contour interferometer (DDCI) studied in \cite{NANOLett2017}. DDCI consists of two superconducting contours. The contours are arranged one above the other and separated by a dielectric in the ideal DDCI shown in Fig.1. The contours  are weakly coupled by Josephson junctions $J_{a}$ and $J_{b}$ in two points, Fig.1. The bias current flows into the upper loop, passes through the Josephson junctions $J_{a}$, $J_{b}$ in the lower loop and flows out the other side of the lower loop, Fig.1.  It was shown theoretically \cite{NANO2010} that the maximum value of the superconducting bias current
$$I _{s} = I _{a}\sin \varphi _{a} + I _{b}\sin (\varphi _{a}+ \pi (n _{u} + n _{d}) \eqno {(6)}$$
through the DDCI (i.e. the critical current $I_{c}$) depends only on the parity of the sum $n _{u} + n _{d}$ of quantum numbers of the upper loop $n _{u}$ and the bottom loop $n _{d}$: $I_{c,e} = I _{a} + I _{b}$ when the sum is even $n _{u} + n _{d} = 2n$ ($I_{c} = 2I _{a}$ at $I _{a} = I _{b}$) and $I_{c,o} = |I _{a} - I _{b}|$ when the sum is odd $n _{u} + n _{d} = 2n+1$ ($I_{c} = 0$ at $I _{a} = I _{b}$). The idea of such a device was inspired by experimental results \cite{Zhilyaev2000} and their explanation \cite{Zhilyaev2001}.  The critical current $I_{c}$ of the DDCI and the voltage at the bias current $I_{c,o} < I_{b} \approx  I_{c,e}$ should jump at a great value when the quantum number change in one of the loops. Therefore the DDCI is an ideal device for observation of the states of the flux qubit. The states can be observed when one of the loops of the DDCI is flux qubit, Fig.1. 

\begin{figure}
\includegraphics{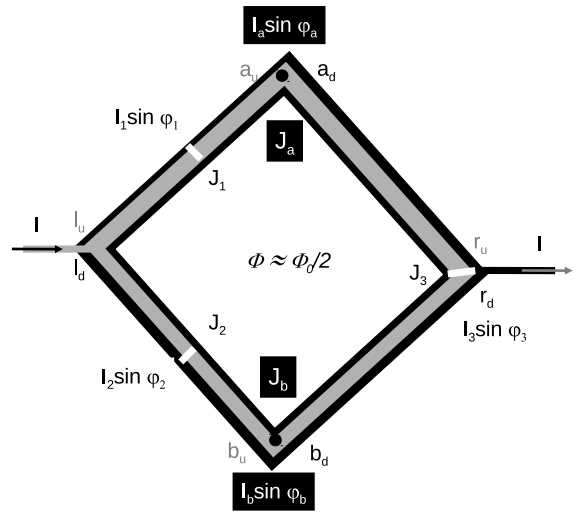}
\caption{\label{fig:epsart} Scheme of the superconducting differential double contour interferometer. The bottom loop (shown in black) is a solid superconducting loop. The upper loop (shown in gray) is the flux qubit with three Josephson junctions $J_{1}$, $J_{2}$, $J_{3}$ (indicated by white rectangles). The upper circuit is separated from the lower loop by a dielectric layer, except for two points (indicated by black circles) in which the phases of the wave function of the two circuits are connected by the Josephson  junctions $J_{a}$, $J_{b}$. The critical current through these junctions is much less than the critical current of the Josephson junctions of the flux qubit  $J_{a}, J_{b} \ll J_{1}, J_{2}, J_{3}$. The measuring bias current flows into the upper loop, passes through the Josephson junctions $J_{a}$, $J_{b}$ in the lower circuit and flows out the other side of the lower circuit. }
\end{figure}  

Measurements have corroborated the voltage jumps when the quantum number in one of the loops change and a bias current $I_{b} > I_{c,o}$ flows through the DDCI \cite{NANOLett2017}. The both superconducting loops of the DDCI used in \cite{NANOLett2017} were not interrupted by the Josephson junctions. Therefore the quantum number changed when the persistent current $I_{p,i} = (n_{i}\Phi_{0} - \Phi )/L_{k}$ in the loop $i$  reached a critical value $I_{c,i} \approx \Phi_{0}l/2\pi \xi (T)L_{k}$ with the magnetic flux $\Phi $ variation \cite{nJump2003,nJump2003} rather than at $\Phi \approx (n'+0.5)\Phi_{0}$. Here $L_{k}$ is the kinetic inductance; $l$ is the length of the loop; $\xi (T)$ is the coherence length of the superconductor. The voltage jumps up when the sum $n _{u} + n _{d}$ becomes odd because of the change the quantum number in one of the loops and returns back when the quantum number changes in another loop \cite{NANOLett2017}. The period $B_{0} = \Phi_{0}/S$ between the jumps up or down corresponds the flux quantum $\Phi_{0}$ inside the loop \cite{NANOLett2017}. More than 1000 jumps with period $B_{0} \approx 0,053 \ Oe$ were observed at low temperatures in the interval $30 \ Oe < B < 30 \ Oe$ at the measurement of the DDCI with the side of the square loops $a \approx 20 \ \mu m$ and the area $S = a^{2} \approx 400 \ \mu m^{2}$ \cite{NANOLett2017}. The superconducting current (6) and the critical current through the DDCI should not depend directly on the magnetic field in the ideal case, when the loops are arranged exactly one above the other, Fig.1. The loops of the DDCI used in \cite{NANOLett2017} were shifted relatively each other on $a_{sh} \approx 0,6 \ \mu m$ because of the simple shadow evaporation technique used for the fabrication of the real device. Therefore the amplitude of the voltage jumps was modulated in the magnetic field with the period $B _{m} \approx 0.8 \ Oe$ corresponding approximately to the mutual shift of the loops $B _{m} \approx \Phi_{0}/aa_{sh}\surd 2$, see Supporting Information of  \cite{NANOLett2017}.

\section{The theory of the DDCI with a flux qubit as one of the contours}   
The voltage jump $|V(n'+1) - V(n')| > 20 \ \mu V$ observed in \cite{NANOLett2017} allows to detect the quantum state of superconducting loop, in particular the flux qubit. The DDCI can be used for observation of the state of the flux qubit when the critical current of the Josephson junctions connecting the two loops is much smaller than the critical current of the Josephson junctions of the flux qubit which is one of the loops of the DDCI $J_{a}, J_{b} \ll J_{1}, J_{2}, J_{3}$, Fig.1. The critical current of the DDCI in this case is the maximum superconducting current
$$I _{s} = I _{a}\sin \varphi _{a} + I _{b}\sin \varphi _{b} \eqno {(7)}$$
through the two Josephson junctions $J _{a}$ and $J _{b}$. The superconducting current is determined by the critical current $I _{a}$, $I _{b}$ and the phase difference $\varphi _{a}$, $\varphi _{b}$ of these Josephson junctions. The relation between $\varphi _{a}$ and $\varphi _{b}$ is determined by the requirement of uniqueness of the wave function in three circuits: 1) $l _{d} - a _{d}- r_{d} - b_{d} - l _{d}$, 2) $l _{u} - J _ {1} - a _{u}- J _ {3} - b_{u} - J _ {2} - l _ {u}$ and 3) $l _{u} - J _ {1} - a _{u} - a _{d} - r_{d} - b_{d} - b_{u} - J _{2} - l _{u}$, Fig.1. According to the first requirement $\oint_{l}dl \nabla \varphi = \varphi_{ld,ad} + \varphi_{ad,rd} + \varphi_{rd,bd} + \varphi_{bd,ld} = 2\pi n_{d}$ the phase difference between the points $a _{d}$ and $d _{d}$ should be equal $\varphi_{ad,rd} + \varphi_{rd,bd} =  \pi n_{d}$ when the lower loop without Josephson junctions is homogeneous in which $\varphi_{ad,rd} + \varphi_{rd,bd} = \varphi_{bd,ld} + \varphi_{ld,ad}$ and the persistent current $I_{p,d}$ in it is much larger than the Josephson critical current $J_{a}, J_{b}$. The influence of the measuring current on the phase change in the lower loop can be neglected when $I_{p,d} \gg J_{a} + J_{b}$. Here $n_{d}$ is the quantum number of the lower loop. The second requirement $\oint_{l}dl \nabla \varphi = \varphi_{lu,J1} + \varphi_{1} + \varphi_{J1,au} + \varphi_{au,J3} + \varphi_{3} + \varphi_{J3,bu} + \varphi_{bu,J2} + \varphi_{2} + \varphi_{J2,lu} = 2\pi n_{q}$ determines the phase difference between points $a _{u}$ and $b_{u}$ of the flux qubit, where $n_{q}$ is the quantum number of the flux qubit. This requirement may be rewritten as $\varphi_{1} + \varphi_{3} + \varphi_{2} + 2\pi\Phi/\Phi_{0} \approx 2\pi n_{q}$ since  $\varphi_{lu,J1} + \varphi_{J1,au} + \varphi_{au,J3} + \varphi_{J3,bu} + \varphi_{bu,J2} + \varphi_{J2,lu}  \approx q\Phi /\hbar = 2\pi\Phi/\Phi_{0}$. The relationship between the phase differences $\varphi _{a}$ and $\varphi _{b}$ can be found from the third requirement $\oint_{l}dl \nabla \varphi = \varphi_{lu,J1} + \varphi_{1} + \varphi_{J1,au} + \varphi_{a} + \varphi_{ad,rd} + \varphi_{rd,bd} - \varphi_{b} + \varphi_{bu,J2} + \varphi_{2} + \varphi_{J2,lu} = 2\pi n$.   

In the equality $\varphi_{b} = \varphi_{a} + \varphi_{1} + \varphi_{2} + (\varphi_{ad,rd} + \varphi_{rd,bd}) + (\varphi_{lu,J1} +  \varphi_{J1,au} +  \varphi_{bu,J2} + \varphi_{J2,lu}) - 2\pi n$ the quantum number can be taken zero $ n =0$ since $\sin (\varphi _{b} - 2\pi n) \equiv \sin \varphi _{b}$ in (7). Then
$$\varphi_{b} = \varphi_{a} + \varphi_{1} + \varphi_{2} + \pi n_{d} + \pi \frac{\Phi }{\Phi _{0}} =$$ 
$$= \varphi_{a} + \varphi_{1} + \varphi_{2} + \pi (n_{d} + n'+0.5+\delta f) \eqno{(8)}$$ 
We take $\Phi /\Phi _{0} = n'+0.5+\delta f$ and $n_{d} = n'$ since the states of the flux qubit should be observed near $\Phi = (n'+0.5)\Phi _{0}$ and when the quantum number of the lower loop $n_{d}$ is constant. In this case the superconducting current through the DDCI 
$$I _{s} = I _{a}\sin  \varphi _{a} + I _{b}\sin  (\varphi _{a}+ \varphi_{1} + \varphi_{2} +0.5\pi + \pi \delta f) \eqno{(9)}$$
depends first of all of the sum of the phase changes $\varphi_{1} + \varphi_{2}$ on the two Josephson junctions of the flux qubit. The sum of the integer numbers is taken zero $ n_{d} + n' = 2n' = 0$ since $\sin (\varphi _{a}+ \varphi_{1} + \varphi_{2} +0.5\pi + \pi \delta f + 2\pi n') \equiv \sin (\varphi _{a}+ \varphi_{1} + \varphi_{2} +0.5\pi + \pi \delta f)$. The sum $\varphi_{1} + \varphi_{2}$ is determined by the equality 
$$\varphi_{1} + \varphi_{2} + \varphi_{3} = 2\pi (n_{q}- \frac{\Phi }{\Phi _{0}}) = 2\pi (n_{q}- n'-0.5-\delta f) \eqno{(10)}$$ 
obtained from the second requirement. According to the equality (10) $\varphi_{1} + \varphi_{2} + \varphi_{3} \approx -\pi$ at $n_{q} = n'$ and $\varphi_{1} + \varphi_{2} + \varphi_{3} \approx +\pi$ at $n_{q} = n'+1$ when $\delta f \ll 1$. The phase differences should have the same sign because of the equality of the current through the Josephson junctions of the flux qubit $I_{p} = I_{1} sin \varphi_{1} = I_{2} sin \varphi_{2} = I_{3} sin \varphi_{3}$. Therefore the critical current $I_{c}$ of the DDCI, i.e. the maximum value of the superconducting current (9) should jump on a big value when the state of the flux qubit changes between $n_{q} = n'$ and $n_{q} = n'+1$: for example, $I_{c} \approx \max |I _{a}\sin  \varphi _{a} + I _{b}\sin  (\varphi _{a}+ \pi /6)|$ at $n_{q} = n'$ and $I_{c} \approx \max |I _{a}\sin  \varphi _{a} + I _{b}\sin  (\varphi _{a}+ 5\pi /6)|$ at $n_{q} = n' + 1$ when $I_{1} = I_{2} = I_{3}$. The jump of the DDCI critical current has maximum value when $I _{a} = I _{b}$ and the ratio between the critical currents $I_{1}$, $I_{2}$, $I_{3}$ of the flux qubit is such that  $\varphi_{1} + \varphi_{2} = \varphi_{3}$: $I_{c} \approx 2I _{a}$  at $n_{q} = n'$ and $I_{c} \approx 0$  at $n_{q} = n'+1$. 

\section{Continuous observation of the state of the flux qubit in time}
Measurements of the switching current of the DC-SQUID, coupling inductively with the flux qubit, give the different values corresponding to the states $n_{q} = n'$ and $n_{q} = n'+1$ with equal probability at $\Phi = (n'+0.5)\Phi_{0}$ \cite{Tanaka2002,Tanaka2002PS}. The results \cite{Tanaka2002,Tanaka2002PS} do not allow to understand why the observed states of the flux qubit change from measurement to measurement. The states can change because of the act of measurement or in time. The method, used in the works \cite{Tanaka2002,Tanaka2002PS,Tanaka2004,Tanaka2009}, cannot be applied for the observation of the change of the states $n_{q} = n'$, $n_{q} = n'+1$ in time since in practice the magnetic coupling of the DC-SQUID with the flux qubit is sufficiently weak. In real qubit measurements, the magnetic flux $\Delta \Phi _{I} = LI_{p}$ induced by the qubit circulating current $I_{p}$ is $\Delta \Phi _{I} = 10^{-3}\Phi_{0} \div 10^{-2}\Phi_{0}$ \cite{Tanaka2004the}. Therefore, the state of the qubit is observed with the help of numerous acts of measurement in which the bias current applied to the dc SQUID increases every time from $I_{b} = 0$ to $I_{b} \approx  I_{c}(\Phi )$ \cite{Tanaka2002,Tanaka2002PS,Tanaka2004,Tanaka2009}. 

The state of the flux qubit can be observed continuously in time with the help of the DDCI thanks to the great jump of the voltage at the change of the quantum number of one of its loops \cite{NANOLett2017}. The voltage jumps $|V(n'+1) -  V(n')| \approx 20 \ \mu V$ are observed \cite{NANOLett2017} when the bias current $I_{b} \approx 0.02 \ \mu A$ is much less than typical values of the persistent current $I_{p} \approx 0.5 \ \mu A$  of the flux qubit measured in \cite{Mooij2000,Tanaka2004,Tanaka2009}. The requirement $J_{a}, J_{b} \ll J_{1}, J_{2}, J_{3}$ will be fulfilled if we will combine the flux qubit measured in the works \cite{Mooij2000,Tanaka2004,Tanaka2009} and the DDCI device investigated in the work \cite{NANOLett2017}. The technology allows to increase the difference between the values $J_{a}, J_{b}$ and $J_{1}, J_{2}, J_{3}$ and, thus, to investigate the influence of the bias current $I_{b}$ on the behaviour of the flux qubit. The critical current of the DDCI, i.e. the maximum value of the superconducting current (9), should depend on the state of the flux qubit when $\delta f \ll 1$: at $n_{q} = n'$ the phase differences are negative $\varphi_{1} + \varphi_{2} < 0$ according to (10) and the critical current has a minimal value $I_{c,min}$ ($I_{c,min} = 0$ in the ideal case) whereas at  $n_{q} = n'+1$ the phase differences are positive $\varphi_{1} + \varphi_{2} > 0$ and the critical current has a maximum value $I_{c,max}$ ($I_{c,max} = 2I_{a}$ in the ideal case). Therefore the voltage jump will be observed when the state of the flux qubit changes and in the case when the bias current exceeds the minimal critical current $I_{b} > I_{c,min}$.   

Any observations of the voltage jumps (or their absence) at a bias current constant in time can be of fundamental importance. The absence of the jumps may mean that the changes of the flux qubit state from measurement to measurement observed in \cite{Tanaka2002,Tanaka2002PS,Tanaka2004,Tanaka2009} occur because of the act of measurement. If the jumps will be observed, then it will be possible to measure the frequency of the switching between the states of the quantum two-level system with the strongly discrete spectrum. Measurements of the frequency spectrum and its dependence on temperature and other parameters can provide important information about such systems. In particular such measurements can give an additional information about thermally activated behavior and macroscopic quantum tunneling \cite{MQT2016PhysRep,MQT2015Nature,MQT2003PRL}. It is not excluded that the observation of the flux qubit state with the help of the DDCI will allow to clarify the nature of the strange $\chi $-shaped crossing of the distribution of the switching current as a function of the applied magnetic flux observed in \cite{Tanaka2004,Tanaka2009}. The results of the single-shot readouts correspond to the two values of the persistent current (1) of the permitted states $n_{q} = n'$, $n_{q} = n'+1$ of the flux qubit at $|\delta \Phi | > 0.002\Phi _{0}$ \cite{Tanaka2004,Tanaka2009}. But the switching current measured at $\delta \Phi = 0$ correspond to the value of the persistent current $I_{p} = 0$ forbidden according to the quantization condition (1).

\section{Using the DDCI with the flux qubit to measure magnetic flux}
The probability (5) changes in a narrow interval of the magnetic flux from 1 at  $\delta f \approx -0.01$ to 0 at  $\delta f \approx 0.01$ \cite{Tanaka2004} due to the strong discreteness of the permitted state spectrum of real superconducting loop. The value $I_{p}\Phi_{0}/k_{B}$ corresponds to the temperature $\approx 100 \ K$ at the persistent current $I_{p} \approx 0.5 \ \mu A$ of the flux qubit investigated in \cite{Tanaka2004} and $\theta = I_{p}\Phi_{0}/k_{B}T \approx 100$ at $T \approx 1 \ K$ \cite{NANOLett2017}. Therefore the average value of the switching current $\overline{I_{sw}} = P(n')I_{sw}(n') + P(n'+1)I_{sw}(n'+1) $ changes in the interval $|\delta f| \approx 0.01$ which is much smaller than the interval of the magnetic flux $|\delta f| \approx 0.5$ in which the critical current of the dc SQUID changes \cite{Barone1982}. The average value of the critical current $\overline{I_{c}} = P(n')I_{c}(n') + P(n'+1)I_{c}(n'+1) $ of the DDCI with a flux qubit as one of the contours should also change in this narrow interval. The voltage 
$$\overline{V} = \Theta ^{-1}\int _{\Theta} dt V(t) \approx V _{min}P(n') + V _{max}P(n'+1) \eqno{(11)} $$ 
averaged in time $\Theta $ should also change in this narrow interval since the jumps of the voltage are observed at the change of the quantum number of the loop, according to the experimental results \cite{NANOLett2017}, when a bias current $I_{b} > I_{c,o}$ flows through the DDCI. The averaging time $\Theta $ should exceed the period between the jumps. We don't know how often the flux qubit will switch between the permitted states $n_{q} = n '$ and $n_{q} = n'+1$. It has to be measured. But it can be expected that the switching frequency can be high in some cases. 

The sensitivity of the dc SQUID is determined by the steepness of the voltage dependence on the magnetic flux, i.e. the gradient $(\partial V/\partial \Phi ) _{I}$ \cite{SQUIDs1977}. The critical current and the voltage of the dc SQUID changes in the interval $\Delta \Phi = \Phi _{0}/2$ according to the relation $I _{c} = 2 I _{c,j}|\cos \pi \Phi |$ \cite{Barone1982} valid in the case of weak screening $\beta = 2L I _{c,j}/\Phi _{0} \ll 1$  \cite{SQUIDs1977}. The voltage change in this interval $\Delta V = R _{d}\Delta I _{c}$ cannot exceed the value $R _{d}I _{c} < \Delta /e$, where $ R _{d}$ is the dynamical resistance of the Josephson junctions, $\Delta $ is the energy gap of the superconductor and $e$ is the electron charge \cite{Barone1982}. Therefore the maximum value $(\partial V/\partial \Phi ) _{I}$ of the classical dc SQUID cannot exceed $2\Delta /e\Phi _{0}$. The real value $(V/\Phi ) _{I} \approx 2 \ \mu V/\Phi _{0}$ \cite{SQUIDs1977} of a typical dc SQUID is substantially smaller than the maximum value. The voltage jumps $V _{max} - V _{min} \approx 20 \ \mu V$ observed in \cite{NANOLett2017} is smaller the maximum possible value but the theory does not exclude that the jump can reach this value. The sensitivity of the DDCI with a flux qubit as one of the contours can significantly exceed the sensitivity of the conventional dc SQUID thanks to the small interval of change in the average voltage (11). For example, the gradient $(\partial \overline{V}/\partial \Phi ) _{I}$ can exceed $ \approx 20 \ mV/\Phi _{0}$ when the voltage jump $V _{max} - V _{min} \approx 20 \ \mu V$ \cite{NANOLett2017} and the interval of the probability $P(n')$ change is equal $|\delta f| \approx 0.001$, as it is observed in \cite{Tanaka2004} at the temperature $T \approx 0.25 \ mK$. This interval can be observed at a higher temperature when the flux qubit with a higher persistent current $I_{p}$ is measured since $|\delta f| \approx 1/\theta = k_{B}T/I_{p}\Phi_{0}$  according to (5). 

The DDCI has the advantage in the measurement of extremely small magnetic field. The sensitivity of the magnetic field $B$ is determined by the sensitivity of the magnetic flux $\Phi $ and the area of the loop $S$ since $B = \Phi /S$. The area $S$ of the dc SQUID cannot be too large because of the strong screening $\Phi _{I} = LI _{c,j}> \Phi _{0}/2$ in the loop with a high magnetic inductance $L \approx \mu _{0}l$ in which $\beta = 2L I _{c,j}/\Phi _{0} > 1$ \cite{SQUIDs1977}. Therefore the flux transformer is used for the measurement of extremely small magnetic fields \cite{SQUIDs1977}. The area $S$ of the DDCI loops can be large since the magnetic flux $\Phi _{I} = LI _{p} = (L/L _{k})(n\Phi_{0} - \Phi) \approx (s/\lambda _{L}^{2}(T))(n\Phi_{0} - \Phi)$ \cite{QuSMF2016} induced by the persistent current $I _{p}$ does not depend on the loop size $l$. Therefore the DDCI can be used for the measurement of extremely small magnetic field without of the  flux transformer.

\section{Conclusion} 
We draw reader's attention on the fundamental and practical importance of the voltage jumps at the change of the quantum number of one of the loops of the  superconducting differential double contour interferometer (DDCI) observed in \cite{NANOLett2017}. This jumps allow to investigate experimentally a dynamic behaviour macroscopic quantum systems such as the flux qubit. These fundamentally new investigates can help to answer questions that arise in connection with the superposition of macroscopic quantum states and quantum tunneling between these states: for example, "How can the superposition of states with macroscopically different angular momentum be possible?" \cite{Nikulov2010FQ}. Questions arise in the connection not only with quantum tunneling, but also with thermal activation and non-equilibrium noise. The magnetic moment $M_{m} = I_{p}S$ and the angular momentum of Cooper pairs $M_{p} = (2m_{e}/e)M_{m}$  change on macroscopic values $\approx N_{s}\mu _{B}$ and $\approx N_{s}\hbar$ at the switching of the quantum states between $n'$ and $n'+1$ in all these cases: for example, the magnetic moment equals approximately $M_{m} \approx 0.5 \ 10^{5} \ \mu _{B}$ and the angular momentum of Cooper pairs equals $M_{p,n'} \approx  0.5 \ 10^{5} \ \hbar $ when the persistent current $I_{p} \approx  5 \ 10^{-7} \ A$ and the area $S \approx 10^{-12} \ m^{2}$ correspond to the typical flux qubit \cite{Mooij2003}. Here $N_{s}$ is the number of Cooper pairs in the loop, $\mu _{B}$ is the Bohr magneton and $\hbar $ is the reduced Planck constant. 

What force can change the angular momentum on the macroscopic value? According to the Ehrenfest theorem \cite{Ehrenfest1927} Newton's second law can be applied to average values of quantum systems. We may write $dM_{p}/dt = rF_{x}$ if the Ehrenfest theorem can be applied to macroscopic quantum system such as superconducting ring with the radius $r$. We know from the experimental results \cite{Tanaka2002,Tanaka2002PS,Tanaka2004,Tanaka2009} that the angular momentum of Cooper pairs in the flux qubits changes on the value $\Delta M_{p} = |M_{p,n'+1} - M_{p,n'}| \approx 10^{5} \ \hbar $. If this change occurs under the influence of a force $F_{x}$, we can estimate the magnitude of this force by observing the dynamics of the state change and measuring the time during which this change occurs. The question of the force changing the angular momentum is particularly relevant in connection with the observations of the dc potential difference $V_{dc}$ on asymmetric superconducting rings with the persistent current $I_{p}$ \cite{PLA2012PV,APL2016}. The persistent current flows against the total electric field in one of the ring halves according to these observations \cite{Physica2019}. 

Quantum theory predicts the jump of the critical current at the change of the quantum number not only of the DDCI loops but also superconducting rings with asymmetric link-up of current leads. The probability of the quantum states $n'$ and $n'+1$ of superconducting rings with asymmetric link-up of current leads should also change in the narrow interval of magnetic flux  $\delta \Phi $ near $\Phi = (n'+0.5)\Phi_{0}$ in accordance with the Arrhenius law (5). The change of the average value of the critical current and the voltage (11) in a small interval of the magnetic flux $\delta \Phi $ was proposed to be used to create a magnetometer with high sensitivity and easy to manufacture \cite{Letters2014}. But the first measurements of aluminium rings with asymmetric link-up of current leads revealed that a smooth change of the critical current is observed near $\Phi = (n'+0.5)\Phi_{0}$ instead of the jump \cite{PLA2017}, contrary to the theoretical prediction. In contrast to these experimental results, the observations of the voltage jumps \cite{NANOLett2017} guarantees the possibility to make a sensitive magnetometer based on the DDCI.

\section*{Acknowledgments}
This work was made in the framework of State Task No 075-00475-19 -00.


\begin{thebibliography}{99}

\bibitem{Leggett1985} A.J. Leggett and A. Garg, Phys. Rev. Lett. {\bf 54}, 857 (1985)
\bibitem{LGineq2014} C. Emary, N. Lambert, and F. Nori, Rep. Prog. Phys. {\bf 77}, 016001 (2014). 
\bibitem{LGineq2016} G.C. Knee, K. Kakuyanagi, M.C. Yeh, Y. Matsuzaki, H. Toida, H. Yamaguchi, S. Saito, A.J. Leggett, W.J. Munro, Nature Comm. {\bf 7} 13253 (2016)
\bibitem{Clarke2008}  J. Clarke and F.K. Wilhelm, Nature {\bf 453}, 1031 (2008).
\bibitem{Devoret2013} H. Devoret and R. J. Schoelkopf, Science {\bf 339}, 1169 (2013).
\bibitem{Wendin2017} G Wendin, Rep. Prog. Phys. {\bf 80}, 106001 (2017). 
\bibitem{China2018} W.-Y. Liu, D.-N. Zheng, and S.-P. Zhao, Chin. Phys. B {\bf 27}, 027401 (2018) 
\bibitem{Mooij1999} J. E. Mooij, T. P. Orlando, L. Levitov, L. Tian, C.H. van der Wal, S. Lloyd, Science {\bf  285}, 1036 (1999).
\bibitem{Mooij2003} I. Chiorescu, Y. Nakamura, C. J. P. M. Harmans, J. E. Mooij, Science {\bf 299}, 1869 (2003).
\bibitem{Clarke2003}  J. Clarke, Science {\bf 299}, 1850 (2003).
\bibitem{Semba2017} F. Yoshihara, T. Fuse, S. Ashhab, K. Kakuyanagi, S. Saito and K. Semba, Nature Phys. {\bf  13}, 44 (2017)
\bibitem{Leggett2003} A. J. Leggett, Science {\bf 296}, 861 (2003).
\bibitem{Leggett2014} M.C. Yeh and A.J. Leggett, arXiv:1401.4186 (2014) 
\bibitem{LL} L.D. Landau and E.M. Lifshitz, Quantum Mechanics: Non-Relativistic Theory, Volume 3, Third Edition, Elsevier Science, Oxford, 1977
\bibitem{Mermin1993} N.D. Mermin, Rev. Mod. Phys. {\bf 65}, 803 (1993). 
\bibitem{Neumann1932} J. von Neumann, Mathematishe Grundlagen der Quantem-mechanik. Springer, Berlin 1932; Mathematical Foundations of Quantum Mechanics, Princeton, NJ: Princeton Univ. Press 1955.
\bibitem{Dirac1930} A.M. Dirac, The Principles of Quantum Mechanics. Oxford University Press, 1958.
\bibitem{Mooij2000} C.H. van der Wal, A.C.J. ter Haar, F.K. Wilhelm, R.N. Schouten, C.J.P.M. Harmans, T. P. Orlando, S. Lloyd, and J.E. Mooij, Science {\bf 290}, 773 (2000).
\bibitem{Tanaka2002} H. Tanaka, Y. Sekine, S. Saito, H. Takayanagi, Physica C {\bf 368}, 300 (2002)
\bibitem{Tanaka2002PS} H. Takayanagi, H. Tanaka, S. Saito, and H. Nakano, Physica Scripta {\bf T102}, 95 (2002)
\bibitem{Tanaka2004} H. Tanaka, S. Saito, H. Nakano, K. Semba, M. Ueda, and H. Takayanagi, arXiv:cond-mat/0407299 (2004).
\bibitem{Tanaka2009} K. Semba J. Johansson K. Kakuyanagi H. Nakano S. Saito H. Tanaka H. Takayanagi, Quantum Inf. Process. {\bf 8}, 199 (2009)
\bibitem{SQUIDs1977} B.B. Schwartz and S. Foner (Eds.) Superconductor Applications: SQUIDs and Machines, Plenum, New York, 1977.
\bibitem{Barone1982} A. Barone and G. Paterno, Physics and Applications of the Josephson Effect. Wiley, New York, 1982.
\bibitem{NANOLett2017} V.L. Gurtovoi, V.N. Antonov, A.V. Nikulov, R.S. Shaikhaidarov, V.A. Tulin, Nano Lett. {\bf 17}, 6516 (2017); arXiv:1710.05728 (2017) 
\bibitem{NANO2010} A.V. Nikulov, Proceedings of 18th International Symposium "NANOSTRUCTURES: Physics and Technology" St Petersburg: Ioffe Institute, p. 367 (2010); arXiv: 1006.5332 (2010) 
\bibitem{Zhilyaev2000} I. N. Zhilyaev, S. G. Boronin, K. Fossheim. Physica C, {\bf 332}, 422 (2000).
\bibitem{Zhilyaev2001} I. N. Zhilyaev, S. G.Boronin, A. V. Nikulov and K. Fossheim, Quantum Computers and Computing, {\bf 2}, 49 (2001).
\bibitem{nJump2003} D.Y. Vodolazov, F. M. Peeters, S. V. Dubonos, A. K. Geim,  Phys. Rev. B {\bf 67}, 054506 (2003)
\bibitem{nJump2007} H. Bluhm, N. C. Koshnick, M. E. Huber, K. A. Moler, arXiv: 0709.1175 (2007). 
\bibitem{Tanaka2004the} H. Nakano, H. Tanaka, S. Saito, K. Semba, H. Takayanagi, and M. Ueda, arXiv: cond-mat/0406622 (2004).
\bibitem{MQT2016PhysRep} J.A. Blackburn, M. Cirillo, N. Gronbech-Jensen, Phys. Rep. {\bf 611}, 1 (2016)
\bibitem{MQT2015Nature} D. Massarotti, A. Pal, G. Rotoli, L. Longobardi, M. G. Blamire, F. Tafuri, Nature Comm. {\bf 6}, 7376 (2015)
\bibitem{MQT2003PRL} F. Balestro, J. Claudon, J. P. Pekola, and O. Buisson, Phys. Rev. Lett. {\bf 91}, 158301, (2003) 
\bibitem{QuSMF2016} A.V. Nikulov, Quant. Stud.: Math. Found. {\bf 3}, 41 (2016) 
\bibitem{Nikulov2010FQ} A.V. Nikulov, Quantum Computers and Computing, {\bf 10}, 42 (2010)
\bibitem{Ehrenfest1927} P. Ehrenfest, Zeitschrift fur Physik. {\bf 45}, 455 (1927). 
\bibitem{PLA2012PV} A.A. Burlakov, V.L. Gurtovoi, A.I. Ilin, A.V. Nikulov, V.A. Tulin, Phys.Lett. A {\bf 376}, 2325 (2012)
\bibitem{APL2016} V.L. Gurtovoi, M. Exarchos, V.N. Antonov, A.V. Nikulov, V.A. Tulin, Appl. Phys. Lett. {\bf 109}, 032602 (2016)
\bibitem{Physica2019} V.L. Gurtovoia, V.N. Antonov, M. Exarchos, A.I. Il'in, A.V. Nikulov, Physica C {\bf 559}, 14 (2019)
\bibitem{Letters2014} A. A. Burlakov, V. L. Gurtovoi, A. I. Il'in, A. V. Nikulov, and V. A. Tulin JETP Letters, {\bf 99}, 169 (2014).
\bibitem{PLA2017} A.A. Burlakov, A.V.Chernykh, V.L.Gurtovoi, A.I.Ilin, G.M.Mikhailov, A.V.Nikulov, V.A.Tulin, Phys. Lett. A {\bf 381}, 2432 (2017).

















\end{thebibliography}
\end{document}